\documentclass[aps,prb,twocolumn]{revtex4-1}

\usepackage{amsmath}
\usepackage{amssymb}
\usepackage{graphicx}
\usepackage{epsfig}
\usepackage{color}

\newcommand{\Av}[1]{\langle {#1} \rangle}

\newcommand{\eps}{\varepsilon}

\newcommand{\sig}{\sigma}
\newcommand{\spinup}{\uparrow}
\newcommand{\spindown}{\downarrow}

\newcommand{\Vect}[1]{{\mathbf{#1}}}

\newcommand{\kv}{\Vect{k}}

\newcommand{\Rv}{\Vect{R}}

\newcommand{\Vpoto}{\upsilon_{0}}

\newcommand{\aR}{a_{\Rv}}
\newcommand{\aRp}{a_{\Rv'}}

\newcommand{\deltRR}{\delta_{\Rv\Rv'}}
\newcommand{\deltLL}{\delta_{LL'}}

\newcommand{\Struct}{S^{a}_{\Rv'L'\Rv L}(\kappa^2)}

\newcommand{\Daz}{D^{a}_{\Rv l}(z)}

\newcommand{\Kink}[1]{K^{a}_{\Rv' L' \Rv L}({#1})}
\newcommand{\Kinkp}[1]{K^{a}_{\Rv' L' \Rv'' L''}({#1})}
\newcommand{\Kinkd}[1]{\dot{K}^{a}_{\Rv' L' \Rv L}({#1})}

\newcommand{\Dpoles}{D_{\mathrm{poles}}}

\newcommand{\gcoh}{\tilde{g}}
\newcommand{\Dcoh}{\tilde{D}}
\newcommand{\geff}{\bar{g}}
\newcommand{\Deff}{\bar{D}}
\newcommand{\DLMAv}[1]{\left\langle{#1}\right\rangle_{\sig}}

\newcommand{\NOS}{\mathcal{N}}

\newcommand{\intBZ}{\frac{1}{\Omega_{\rm BZ}}\int\limits_{\rm BZ}}

\newcommand{\mc}{\multicolumn}
\graphicspath{{figs/}}

\begin{document}

\title{Self-consistent supercell approach to alloys with local environment
effects}

\author{O.~E.~Peil}
\affiliation{I. Institut f\"{u}r Theoretische Physik, Hamburg University,
Hamburg, Germany}

\author{A.~V.~Ruban}
\affiliation{Royal Institute of Technology, Department of Material Science \&
Engineering, SE-10044, Stockholm, Sweden}

\author{B.~Johansson}
\affiliation{Royal
Institute of Technology, Department of Material Science \& Engineering,
SE-10044, Stockholm, Sweden}
\affiliation{Condensed Matter Theory Group, Department of Physics and Materials
Science, Uppsala University, SE-75121 Uppsala, Sweden}

\begin{abstract}
We present an efficient and accurate method for calculating electronic
structure and related properties of random alloys with a proper treatment
of local environment effects. The method is a generalization of the locally
self-consistent Green's function (LSGF) technique for the exact muffin-tin
orbital (EMTO) method. An alloy system in the calculations is represented
by a supercell with a certain set of atomic distribution correlation functions.
The Green's function for each atom in the supercell is obtained by embedding
the cluster of neighboring atoms lying within a local interaction zone (LIZ)
into an effective medium and solving the cluster Dyson equation exactly.
The key ingredients of the method are locality, which makes it
linearly scaling with the number of atoms in the supercell, and
coherent-potential self-consistency of the effective medium, which
results in a fast
convergence of the electronic structure with respect to the LIZ size.
To test the performance and accuracy of the method,
we apply it to two systems: Fe-rich bcc-FeCr random alloy with and
without a short-range order, and a Cr-impurity on the Fe surface.
Both cases demonstrate the importance of taking into account the
local environment effects for correct description of magnetic
and bulk properties.
\end{abstract}

\maketitle

%
%

\section{Introduction}

%
%
Accurate treatment of the electronic structure of disordered systems is a
highly non-trivial problem, which requires the use of a proper statistical
model. In the case of \emph{metallic} random alloys on an ideal crystalline
lattice, the simplest statistical averaging can be done for a single site
leading to the translationally invariant effective medium best given by the
so-called coherent-potential approximation (CPA).\cite{soven67,taylor67}
The CPA constitutes the
basis of most of the first-principles techniques for the electronic-structure
calculations of random alloys, where it is usually combined with the multiple
scattering theory, or Korringa-Kohn-Rostocker
(KKR),\cite{korringa47,kohn54,gyorffy72} and related
methods.\cite{abrikosov91,kudrnovsky91,vitos01} Although the CPA has proven to
be a quite successful approach for the electronic structure of many alloy
systems, its application range is restricted either to completely random alloys
or/and to alloys where local environment effects in the electronic structure
are small, in the sense that their average is accurately given by the
corresponding CPA effective medium.

%
%
The proper account of the local environment effects in the electronic
structure calculations, brought about either by correlated atomic distribution of
alloy components (atomic short range order) or by multisite electronic
structure correlations, can be done in several ways which might be loosely
subdivided into analytical Green's-function-based approaches, in which
the non-locality is taken into account by the summation of some subset of
diagrams, and into supercell methods that make use of the self-averaging
property of certain quantities, implying that the configurational averaging
can be replaced with the averaging over a sufficiently large supercell.

%
%
The main objective of analytical approaches is to obtain a full Green's
function or, alternatively, a self-energy of a disordered system.
This is achieved by summing a large subset (ideally all) of diagrams
corresponding to multiple
scattering from different sites either for the whole system, like it is
done in the augmented space formalism,\cite{mookerjee73, *mookerjee73b}
or for a specific set
of clusters representing a given alloy system, as it is implemented
in the cluster extensions of CPA.\cite{tsukada69, mills78, rowlands03}
The most developed example of this family of methods is the non-local CPA,
that has been recently combined with
first-principles calculations.\cite{rowlands03,biava05,rowlands08}
In this technique, the averaging is
performed over all possible configurations of a specifically chosen cluster
which tiles the entire underlying crystal lattice. In this case, as in
similar analytical approaches, the average translational symmetry
of the underlying lattice is lost together with the corresponding
simple reciprocal space formalism, although it can be recovered
with some additional effort.\cite{rowlands08} 
However, to our knowledge, there has been no implementation
of the non-local CPA within density functional theory (DFT) 
strictly consistent with the total charge density.

%
%
The requirement of the charge self-consistency is important not only
for accurate total-energy calculations,
but also because the correct account of the local environment
effects themselves demands taking
into consideration the response of the electronic density to the
whole surrounding system, which can be fulfilled if the one-electron potential
of every atom is determined in the DFT-self-consistent way.
The difficulty of achieving this self-consistency in Hamiltonian-based
approaches, such as augmented space method, seems to be the main
hurdle on the way to establish them as an accurate tool for
the first-principles calculations of disordered alloys.

%
%
One of the ways to treat a disordered system within DFT self-consistently
is to replace the configurational averaging by the averaging over
a large sample, which is at the heart of supercell-based methods.
These averagings are equivalent for an infinite-size
sample and for quantities that possess the property of self-averaging,
i.e. almost all observables of interest, including
the Green's function and total energy. Moreover, if the interatomic
interactions in the system are screened and hence short-ranged, as it is
the case for most metallic alloys, the size of the sample can be
chosen to be reasonably small without introducing a significant
error.
\footnote{Let us note that this is not the case for non-metallic
systems, where
the effective short-range screening is absent. Such systems require
the use of special techniques\cite{dagram97,wang98,boykin07,popescu10}
and very large supercells exactly for the reason that all type of
interactions are extremely long-range and strong. At the same time,
the methods based on the CPA do not work in this case, since the
CPA cannot provide accurate description of the state in the
band gap and close to the band edge, which are the most important
from the point of view of the majority of the physical properties.}

%
%
In a na\"ive, or direct, implementation, one would set up a supercell
with periodic
boundary conditions, solve it exactly by, for instance, evaluating
the Green's function of the entire supercell within a DFT code,
and then average the on-site components of the Green's function for
every component to get the alloy Green's function. This approach
suffers from at least three serious problems: first, it can be computationally
extremely demanding to find a Green's function of a large supercell,
since the complexity usually scales as the third power of the supercell volume;
second, the Bloch states of a periodic system, with their infinite lifetimes,
cannot properly represent the states of a disordered alloy, and even after
averaging over sites, the alloy Green's function would "feel" the
periodicity, which may result, for example, in an ill-defined conductivity;
and finally, the Brillouin zone (BZ) of the underlying lattice is
shrunk to a small BZ of the supercell, and additional
approximate routines are
needed to restore the original translational symmetry.
Note, that the problem with Bloch states can be partially
resolved by performing additional averaging over different
realizations of the supercell, but this approach is obviously
computationally expensive.

%
%
A method that overcomes all of the named problems is the
locally self-consistent Green's function (LSGF)
technique.\cite{abrikosov96,abrikosov97}
The method, as it will be described below, combines the best features
of both the analytical and supercell-based approaches and in addition,
it is manifestly charge-self-consistent.
In a historical perspective, it can be considered as a generalization
and extension of the multishell, or embedded-cluster method
(ECM),\cite{gonis77} and the
locally self-consistent multiple-scattering (LSMS) method.\cite{wang95}

%
%
The main idea of LSGF is that the local (on-site) Green's function
for each atom in the supercell
is determined from the Dyson equation restricted to the local interaction zone (LIZ)
consisting of the given atom and its local environment
embedded in the CPA-like effective
medium which is, in turn, built upon all the atoms residing on the alloy
(sub)lattice. In this case, the periodic boundary conditions for the
whole supercell are used
only to determine the local environment of atoms close to the boundary
of the supercell and for solving the Poisson equation in the electrostatic
problem. The translational symmetry of the \emph{underlying} lattice
is used for solving the CPA equation for the effective medium
and this method, thus, allows
the use of the proper reciprocal-lattice formalism, avoiding at the same
time the ideal non-decaying Bloch states in the case of alloys, which plague
the direct supercell approach. Besides the mentioned advantages,
the computational complexity of LSGF scales linearly with the supercell
size.

%
%
Certainly, the main playground of the LSGF method is the consistent and
accurate calculations of the electronic structure of random alloys
with or without short range order.
Also, increasing the size of the LIZ allows the systematic
investigation of the local environment effects in alloy systems.
Besides, it is a powerful tool for evaluating screened Coulomb interactions
in random alloys.\cite{ruban02a,ruban02b}
At the same time, the application range of the method reaches far beyond
homogeneous disordered systems.
For example, by judiciously choosing the size of the LIZ to
hide periodic images in a slab geometry, one can efficiently treat
problems involving impurities on or near surfaces. In a similar
way, the explicit evaluation of the interatomic interactions between
a pair or multiple of impurities in the host metal is possible.
The Green's-function approach embodied in the method
provides additional benefits in treating paramagnetic random systems.
Some examples of such applications will be presented in this paper.

%
%
The previous practical realization of the LSGF method was done using
the KKR method within atomic sphere approximation
(ASA).\cite{abrikosov97,ruban99}
The KKR-ASA method itself suffers from the normalization error: electronic states
are normalized within Wigner-Seitz spheres instead of Wigner-Seitz cells.
This leads, in turn, to a quite substantial error in the
total energy, making, for instance, practically impossible the accurate
calculations of the total-energy variations related to the deformation of
the crystal structure.
At the same time, the LSGF approach is a general formalism that can be
implemented in
any code based on multiple-scattering theory. In this paper, we describe
the implementation within the exact-muffin-tin-orbital (EMTO)
method.\cite{andersen94,vitos00,vitos01b,vitos07}
The method belongs to the family of screened-KKR techniques with the basis set
formed by so-called third-generation muffin-tin (MT) orbitals introduced
by Andersen.\cite{andersen94} During the last decade, this method, combined with
the full-charged-density technique, has proven to be sufficiently
accurate in calculations of a wide spectrum of alloy properties.\cite{vitos07}

%
%
The paper consists of two main parts. In the first part, namely
Section~\ref{section:method}, we describe the LSGF formalism, starting
with some details of the EMTO method that are essential for
understanding the implementation. In the second part,
Section~\ref{section:results}, we provide the results
of test calculations followed by real calculations demonstrating
the ability of the method to treat both homogeneous random alloys,
magnetic, as well as non-magnetic, and inhomogeneous systems, such as
surfaces. In the last section, we conclude the results and
briefly discuss possible extensions of the current LSGF implementation.

%
%
\section{Method \label{section:method}}

%
%
As has been mentioned in the Introduction, in LSGF, one
starts from building a supercell with periodic boundary conditions
by replicating a unit cell of the underlying lattice.
The supercell is populated with components in any desired (random or ordered)
configuration (see Subsection~\ref{subsec:sc}). The main
ingredients of the EMTO-LSGF (ELSGF) approach are the
supercell setup,
EMTO method itself, and LSGF part involving the solution of
the restricted Dyson equation.

The EMTO part (more generally, the KKR part) of the ELSGF method
runs similar to the usual EMTO-CPA implementation, as far as
it concerns the effective medium, density of states, charge
density, and total energy. The full symmetry of the underlying
lattice is employed, reflecting the single-site character of
the effective medium. After the effective medium is found, the
cluster path operator is evaluated using the ECM, or restricted
Dyson equation, for each atom in the supercell. This cluster path
operator is subsequently used for the normalization of states,
determining the density of states, charge density, and total energy.

%
%
In this section, we first briefly outline the EMTO Green's-function
formalism and the features specific to its implementation within
the LSGF technique. Then, we present strategies for choosing a proper
supercell, and finally describe the LSGF formalism within EMTO.
Details concerning the structure constants,
the construction of the optimized one-electron potential, the full
charge density, as well as the total energy calculations within the
full-charge-density formalism, are the same as in the usual
EMTO method, and their comprehensive description can be found
in Ref.~\onlinecite{vitos01b, vitos07}.

\subsection{The EMTO Green's-function formalism}

%
%
The main idea behind the EMTO method is to keep the simplicity
of the MT- or screened-KKR method by making use of the muffin-tin geometry
of the one-electron potential, but to improve the accuracy of the electronic
structure calculations to the level of the full potential methods.
The latter can be achieved by replacing usual MT-spheres with large
overlapping MT-spheres which allow a better approximation for the
full potential. At the same time, additional non-overlapping screening
spheres are used to define boundary conditions for the solutions
(referred to as screened spherical waves) in the interstitial region.

%
%
The path operator, $g(z)$, is defined as follows,
\begin{equation}
\sum_{\Rv'' L''}\Kinkp{z}g_{\Rv'' L'' \Rv L}(z) = \deltRR\deltLL,
\end{equation}
with the kink operator
\begin{equation}
\Kink{z} = \aRp \Struct - \deltRR\deltLL \aR \Daz,
\end{equation}
where $\aR$ is the screening-sphere radius at site $\Rv$, $\Struct$ the
screened structure constants depending on energy $\kappa^2 = z - \Vpoto$
defined with respect to the muffin-tin zero $\Vpoto$, and $\Daz$ the
potential function determined as usual by the logarithmic derivative
of the partial waves at the MT sphere. Note the sign convention for
the potential function and structure constants.

For systems with translational symmetry, the on-site path operator is
determined as
\begin{equation}
g_{0}(z) = \intBZ \frac{d\kv}{S(\kv, z) - D(z)},
\end{equation}
where integration is performed over the Brillouin zone of the crystal lattice.
We write all expressions for a one-atom Bravais lattice for clarity; the
extension to multiple-atom basis is straightforward.

To evaluate the density of states (DOS), the path operator must be properly normalized
with the overlap matrix. The overlap matrix in the EMTO formalism is given
by $\Kinkd{z}$, the energy derivative of the kink operator. The number
of states is then,
\begin{align}
\NOS(\eps) = & -\frac{1}{\pi} \int\limits_{C_{\eps}} dz \; \left[G(z) -
\Dpoles(z) \right], \\
G(z) = & \intBZ d\kv\; \frac{\dot{S}(\kv, z) - \dot{D}(z)}{S(\kv, z) - D(z)},
\end{align}
where the integration along the half of the contour embracing the valence band
below energy $\eps$ is performed, and pole contributions, $\Dpoles$,
coming from the poles of $1/D(z)$ and $\dot{D}(z)$, are subtracted. The
Fermi energy, $\eps_F$, is found from the condition $\NOS(\eps_F) = N_{el}$,
where $N_{el}$ is the number of the valence electrons.

%
%
In the case of a random alloy on a lattice, the CPA equations are used
to determine the electronic structure given by the
coherent path operator, $\gcoh$, through the corresponding coherent potential
operator, $\Dcoh$, of the single-site effective medium:
\begin{equation}
\gcoh(z) = \intBZ \frac{d\kv}{S(\kv, z) - \Dcoh(z)}.
\label{eq:gcoh}
\end{equation}

The path operators of the $i$-th alloy component, $g_{i}$, are found via the
single-site Dyson equation,
\begin{equation}
g_{i} = \gcoh + \gcoh (\Dcoh - D_{i}) g_{i},
\label{eq:g_i}
\end{equation}
from which the coherent path operator is determined as
\begin{equation}
\sum_{i} c_{i} g_{i} = \gcoh,
\label{eq:CPA_cond}
\end{equation}
where $c_{i}$ is the concentration of alloy components. The last
three CPA non-linear equations are solved self-consistently.

In EMTO-CPA, the correctly normalized Green's function
and number of states (per Wigner-Seitz cell) are determined as
\begin{align}
\NOS(\eps) = & -\frac{1}{\pi} \int\limits_{C_{\eps}} dz \; \left[ G(z) -
\Dpoles(z) \right], \label{eq:emto_dos1} \\
G(z) = & \intBZ d\kv\; \frac{\dot{S}(\kv, z) - \sum_{i} c_{i} \dot{D}_{i}(z)}
{S(\kv, z) - \Dcoh(z)},
\label{eq:emto_dos2}
\end{align}
and the pole contributions are weighted by concentrations $c_{i}$
accordingly.

\subsection{Supercell\label{subsec:sc}}

%
%
In the LSGF calculations, an alloy system is represented by a supercell model.
In general, creating a supercell with needed statistical properties, given
by its atomic-distribution correlation functions, is a highly non-trivial
task, mathematically equivalent to the optimization of
a many-variable function in a multidimensional space. The initial
building block of the supercell is determined by the underlying lattice
containing $N_{q}$ basis atoms. Let us note that the choice of the initial
unit cell of the supercell is quite arbitrary, and such a unit cell can be different from
that of the underlying crystal lattice in the subsequent LSGF calculations,
provided that the supercell is conformal to the underlying lattice.
A simple example is the choice of the cubic unit cell containing two atoms
in the case of bcc structure as a building block of the supercell and
the use of the bcc translational symmetry in the LSGF calculations.
In general, the choice of the unit cell is motivated by the model
of an alloy system that may have several different sublattices with
different compositions and distributions of alloying elements (for an
application of ELSGF to a rather complex example of the FeCr $\sigma$-phase
with 30 atoms per unit cell subdivided into 5 nonequivalent sublattices
with distinct compositions, see \cite{kabliman11}).

%
%
The setup of the supercell starts from the definition of a desired alloy
configuration by characterizing its atomic correlation functions
that can be defined in different ways. For a homogeneous binary
alloy, for instance, they can be given by the average products of spin-like variables,
$\sigma_i$, taking on values $+1$ or $-1$, depending on which alloy component
occupies site $i$:
\begin{equation} \label{eq:corr_f}
\xi^{(n)}_{f} = \Av{\sigma_{i} \sigma_{j} \dots \sigma_k}_f
\end{equation}
where $\xi^{(n)}_{f}$ is the $n$-site correlation function for cluster
$f$, and $\Av{\dots}$ is the average over the supercell. A completely
disordered configuration is given by
$\xi^{(n)}_{f} = \sigma^n$, with $\sigma = 2c-1 \equiv\xi^{(1)}$, where
$c$ is the concentration of one of the alloy components.

%
%
Although this definition is easily generalized to the case of multicomponent
and inhomogeneous random alloys, where different sublattices with
different alloy compositions and atomic short range order are present, it is extremely
difficult to use it in practice. The main obstacle here is the finite and
quite restricted size of the supercell. For instance, the number of sites
in the supercell, $N$, defines the possible concentrations to be only $k/N$,
where $k = 0,1, \dots N$. Much more severe restrictions to the possible pair
and multi-correlation functions originate from the geometry of the underlying
lattice, e.g., from the number of possible clusters (coordination
number for pair correlation functions). Nevertheless, a
supercell consisting of about 500-2000 sites is usually sufficiently big
to model a large variety of alloy systems, especially taking into
consideration our restricted
knowledge about atomic distribution correlation functions in real alloys.

%
%
The main condition for a supercell to be a valid
representation of a random alloy with a specific short range order is
to have the same correlation functions as the given alloy for those clusters,
$f$, (or coordination shells in case of pair correlation functions)
which affect a physical property of interest. If the cluster expansion
is applicable to a particular observable, $A$, then the {\it relevant} clusters
are those, for which expansion coefficients, $A_f$, have non-zero (in practice
non-negligible) values:
\begin{equation}
A = \sum_f A_f \xi_f.
\end{equation}

Here, we do not discuss for which physical properties such an expansion
can be valid in general, and how fast the expansion converges,
but as an example we can mention the total energy of an alloy which, as usually tacitly assumed, can be expanded in this way. The coefficients $A_f$ are then just the effective interactions of the corresponding Ising Hamiltonian.

Let us note that a finite supercell of a restricted size
cannot represent a random alloy in general
(and called "random" for that matter): the same supercell
can be "random" for the same system for
one property and "ordered" for the other, not mentioning different
systems.
In principle, one should check the configurational dependence of
the observable of interest (or ideally to find out coefficients $A_f$)
prior to using a supercell as a random-alloy model.
For instance, the expansion coefficients of the total energy are related to the effective interatomic interactions that can be found either with the aid of the generalized perturbation method within single-site CPA or with the cluster inversion method.

%
%
In LSGF, the \emph{real} correlations of the supercell are only taken into consideration inside the LIZ, while outside the LIZ, the correlation functions correspond to a completely random alloy. An observable
calculated by the LSGF method is given by the following formula:
\begin{equation} \label{eq:a_lsgf}
A^{LSGF} = \sum_{f \in LIZ} A_f \xi_f + \sum_{f \notin LIZ} A_f \xi_f^{rand} ,
\end{equation}
where $\xi_f^{rand} \equiv \xi_{f}^{(n)-rand} = \sigma^n$ are the correlation
functions of the completely random alloy. The condition for the cluster
to belong to the LIZ is that one of its vertices coincides with the position of the central atom, and all cluster atoms belong to the LIZ. This condition will be elaborated in the next sections, where we give the details of the electronic structure calculations with the LSGF method.

%
%
The electrostatic energy is calculated "exactly" for a given supercell.
This means that the summation is not restricted to the LIZ but is rather
performed over the entire periodic infinite system. In principle, this may result in a spurious electrostatic interaction between periodic images. Fortunately, in most metallic alloys, the pair interactions are screened, the screening length being rather small. The important condition that the range of the pair-correlation function and the screening length are within a supercell can, thus, be easily satisfied for supercells containing several hundred atoms.

\subsection{ELSGF}

%
%
Given a supercell containing $N$ atoms, the calculation of the electronic structure within the ELSGF method starts from the determination of the translationally invariant CPA effective medium built upon all the atoms on a corresponding (sub)lattice. In the simplest case of a Bravais lattice,
corresponding CPA equations are similar to
Eqs.~(\ref{eq:gcoh})-(\ref{eq:CPA_cond}):
\begin{align}
\geff_{0} = & \intBZ d\kv
\frac{1}{S(\kv, z) - \Deff(z)}, \label{eq:lsgf1}\\
g_{i} = & \geff_{0} + \geff_{0} (\Deff_{0} - D_{i}) g_{i}, \label{eq:lsgf2} \\
\geff_{0} = & \frac{1}{N}\sum_{i} g_{i}, \label{eq:lsgf3}
\end{align}
Here, $\geff_{0}$ and  $\Deff$ are the on-site effective-medium path
operator and logarithmic derivative; $g_{i}$ is the on-site path operator of
site $i$. These nonlinear equations are solved self-consistently for $\Deff$
and $\geff$ for a given set of one-electron potentials in the supercell.
Eqs.~(\ref{eq:lsgf1})-(\ref{eq:lsgf3}) are the CPA equations for an $N$-component
alloy. This makes the effective medium single-site-self-consistent
and all translational and point symmetries of the
underlying lattice are preserved.
In addition, a connection to the CPA guarantees the analyticity
of the effective medium and hence of its real-space path operator
determined as
\begin{equation}
\geff_{ij} =  \intBZ d\kv
\frac{e^{i\kv(\Rv_i-\Rv_j)}}{S(\kv, z) - \Deff(z)}, \label{eq:g_clust}
\end{equation}
where $\Rv_i$, $\Rv_j$ are the positions of sites $i$ and $j$ of the lattice.

%
%
Once the effective medium is defined, the electronic structure for every site can be determined by solving the multiple-scattering problem for the LIZ-cluster embedded into the effective medium, which enables one to take the local environment effects into consideration. The size of the LIZ is usually defined as a number of coordination shells constituting the cluster centered around a given atom, with the LIZ of size one (LIZ=1) corresponding to the cluster consisting of a single central atom. The rest of the system outside the LIZ is given by the effective medium.

The multiple-scattering problem for such a setup is solved \emph{exactly} with the aid of the Dyson equation as it is done in the Embedded Cluster Method (ECM).\cite{gonis77} Consider a LIZ-cluster given by atoms with potential functions $D_{i}$, where $i$ runs over cluster sites $\Rv_i$. The cluster is embedded into the effective medium defined by a real-space path operator $\geff_{ij}$ and a coherent potential function $\Deff_{i}$. The path operator of the cluster, $g_{ij}$, is then found within the ECM as
\begin{equation}
g_{ij} = \geff_{ij} + \sum_{k}\geff_{ik} (\Deff_{k} - D_{j}) g_{kj},
\label{eq:ecm}
\end{equation}
from which one finds immediately
\begin{equation}
g_{ij} = \sum_{k}\left[1 - \sum_{k'}\geff_{ik'} (\Deff_{k'} - D_{k})\right]^{-1}
\geff_{kj}.
\end{equation}

%
%
The Fermi energy of the system, $\eps_{F}$, is determined from the normalization condition, $\NOS(\eps_{F})= N_{el}$, for the number of electron states in the supercell, $N_{el}$, where the number of states is defined in a way similar to that in the EMTO method:
\begin{equation}
\NOS(\eps) =  -\frac{1}{\pi} \int\limits_{C_{\eps}} dz\; \left[ G(z) -
\Dpoles(z) \right] ,
\end{equation}
with the Green's function $G(z)$ having an additional contribution compared to that of the EMTO-CPA Green's function (Eq.~\eqref{eq:emto_dos2}). The point is that unlike the EMTO-CPA, where the states are normalized within a single-site Wigner-Seitz cell,
the states in ELSGF are normalized within the entire LIZ.
Starting from the expression for the Green's function in EMTO-CPA
\eqref{eq:emto_dos2}, rewriting it in the real space, and replacing
the effective-medium path operator by the cluster path operator
within the LIZ, one gets
\begin{align}
G(z) = & \frac{1}{N} \sum_{i} \left[ \intBZ d\kv\; \frac{\dot{S}(\kv, z) -
\dot{D}_{i}(z)} {S(\kv, z) - \Deff(z)} + \right. \notag \\
& \left. \sum_{j} (g_{ij} - \geff_{ij})\dot{S}_{ji} \right].
\end{align}

%
The electron density for each site is obtained from the on-site path
operator, $g_{i} \equiv g_{ii}$, in the same way as it is done in the EMTO method.\cite{vitos07} The DFT self-consistency loop is then closed by evaluating the one-electron potential for every site, assuming the
translational symmetry (periodic boundary conditions)
for the entire supercell in order to solve the
electrostatic problem exactly for each site. Finally, after
the self-consistency is reached, the full charge density is determined
in order to perform accurate calculations of the total energy of the
supercell.\cite{vitos01b,vitos07}

\subsection{Disordered-local-moment model for the LIZ}

%
%
Accurate calculations of the electronic structure and energetics of
paramagnetic alloys with the finite magnitudes of local magnetic moments
on atoms present a challenge for modern first-principles methods.
As has been proven in Ref.~\onlinecite{gyorffy85}, if the magnitude of
the local magnetic moments does not fluctuate strongly and the spin-orbit coupling
is negligible, such a state is accurately described by the collinear
disordered-local-moment (DLM) model, where atoms with spin-up
and spin-down orientations of their local magnetic moment are
distributed randomly on the underlying lattice.

%
%
Although the DLM configuration can be modeled by a supercell with
randomly distributed atoms having different spin orientations,
such a na\"ive supercell representation can, in fact, lead to incorrect
results because in reality, magnetic degrees of freedom fluctuate
(transverse fluctuations are implied here) rapidly
and create, thus, a local environment different from the one with static
magnetic moments. In this sense, a CPA-based scheme seems to be a better method to calculate the systems in the DLM state. Besides, a supercell model
with randomly distributed static moments becomes too cumbersome, since specific atomic correlations should be set up not only between alloy components, but also between their spin-up and spin-down counterparts.

%
%
Within the LSGF method, a straightforward DLM-CPA implementation is possible only in the single-site mode (LIZ=1), when the required CPA averaging and thus the Dyson equation for every site and spin can be solved for the appropriately spin-averaged effective medium. However, this simple scheme obviously breaks down when the nearest-neighbor atoms are included in the LIZ. The problem here is that the potential functions of the neighbors of the central atom would incorrectly correspond to a specific magnetic configuration rather than the random one, as it should be in the DLM state.

%
%
Clearly, a correct description of the DLM state implies here that the central atom "sees" its neighbors inside the LIZ in spin-averaged states. Such a solution can be efficiently implemented in the LSGF method by performing partial constrained averaging of the spin states for all the sites inside the LIZ except the central one. This amounts to choosing an appropriate effective potential of a given atom in the same way as it is usually done within CPA-DLM. To be more specific, we start by defining the on-site path operators in two spin-channels by solving corresponding single-site Dyson equations,
\begin{align}
g_{i}^{\spinup} = & \left[1 - \geff_{0}(\Deff - D_{i}^{\spinup})\right]^{-1}\geff_{0}, \\
g_{i}^{\spindown} = & \left[1 - \geff_{0}(\Deff - D_{i}^{\spindown})\right]^{-1}\geff_{0}.
\end{align}

A partially averaged (DLM-averaged) path operator is then introduced for each site
and global spin channel $\sigma$ as
\begin{align}
\DLMAv{g_{i}} = & \frac{1}{2}(g_{i}^{\spinup} + g_{i}^{\spindown}).
\end{align}

The paramagnetic effective medium is now determined from these DLM-averaged
path operators, with the self-consistency condition being
\begin{align}
\geff_{0} = & \frac{1}{N}\sum_{i} \DLMAv{g_{i}}.
\end{align}

Once the effective medium is determined, one can use the following equation
\begin{align}
\DLMAv{g_{i}} = & \geff_{0} + \geff_{0} (\Deff - \DLMAv{D_{i}})\DLMAv{g_{i}},
\end{align}
to find a corresponding DLM-averaged potential function for each site,
\begin{align}
\DLMAv{D_{i}} = & \Deff - \geff_{0}^{-1} + \DLMAv{g_{i}}^{-1}.
\end{align}

We then solve Eq.~\eqref{eq:ecm} for each site assuming that all the atoms in the LIZ surrounding the central one are in the paramagnetic state. The potential function for each non-central atom in the cluster is thus replaced by a DLM-averaged potential function. Specifically, for a given cluster with a central site $i$, the cluster potential function $D_{j}$ is defined as follows,
\begin{align}
D_{j}^{\sig} = \begin{cases}
D_{i}^{\sig}, & j = i, \\
\DLMAv{D_{j}}, & \textrm{otherwise}.
\end{cases}
\end{align}

%
%
\section{Tests and results \label{section:results}}

%
%
A basic test for the ELSGF method is the convergence of quantities of interest (the total energy in the first place) with respect to the size of the LIZ. By construction, the method observes two limits: $\mathrm{LIZ}\to\infty$ ($N\to\infty$), corresponding to the formally exact solution of the Dyson equation for the entire system; $\mathrm{LIZ} = 1$, equivalent to the CPA with the correct account of electrostatics (sometimes referred to as the isomorphous CPA). In between these two limiting cases, the convergence with respect to the LIZ depends pretty much on the observable or, more strictly, on how fast the cluster expansion coefficients decay with distance, as has already been emphasized in Sec.~\ref{subsec:sc}. In particular, the expansion coefficients of the total energy are related to effective interactions, and the convergence test can be used as a rough estimate of the range of the effective interactions.\cite{abrikosov97} An example of such a calculation is
given below.

%
%
Later in this section, we demonstrate some of the capabilities of the ELSGF method by applying it to real systems. Emphasis is made on the effects of short-range order, especially in magnetic systems. Also, a surface-segregation problem is considered as an example of an inhomogeneous system.

\subsection{The range of the relevant correlation functions from ELSGF}

%
%
The direct relation between the convergence with the LIZ size and the range
of the correlation functions makes it possible to estimate the latter
by varying the size of the LIZ in LSGF calculations.
As has been discussed in Sec.~\ref{subsec:sc}, if an observable is
self-averaging and can be expanded
in terms of the atomic-distribution correlation functions,
the expansion is given by Eq.~\eqref{eq:a_lsgf} within LSGF.
According to this equation, the correlation functions beyond the LIZ correspond to
those of a completely disordered alloy, and thus, by calculating a completely ordered
alloy with LSGF, one captures only contributions from the correlation functions
corresponding to the LIZ-cluster. In view of this, one can estimate the range of
the relevant correlations as the minimal size of the LIZ that provides the same
result as the one given by a direct \emph{ab initio}
calculation for the given ordered structure.

\begin{figure}
\begin{center}
\includegraphics[width=\linewidth]{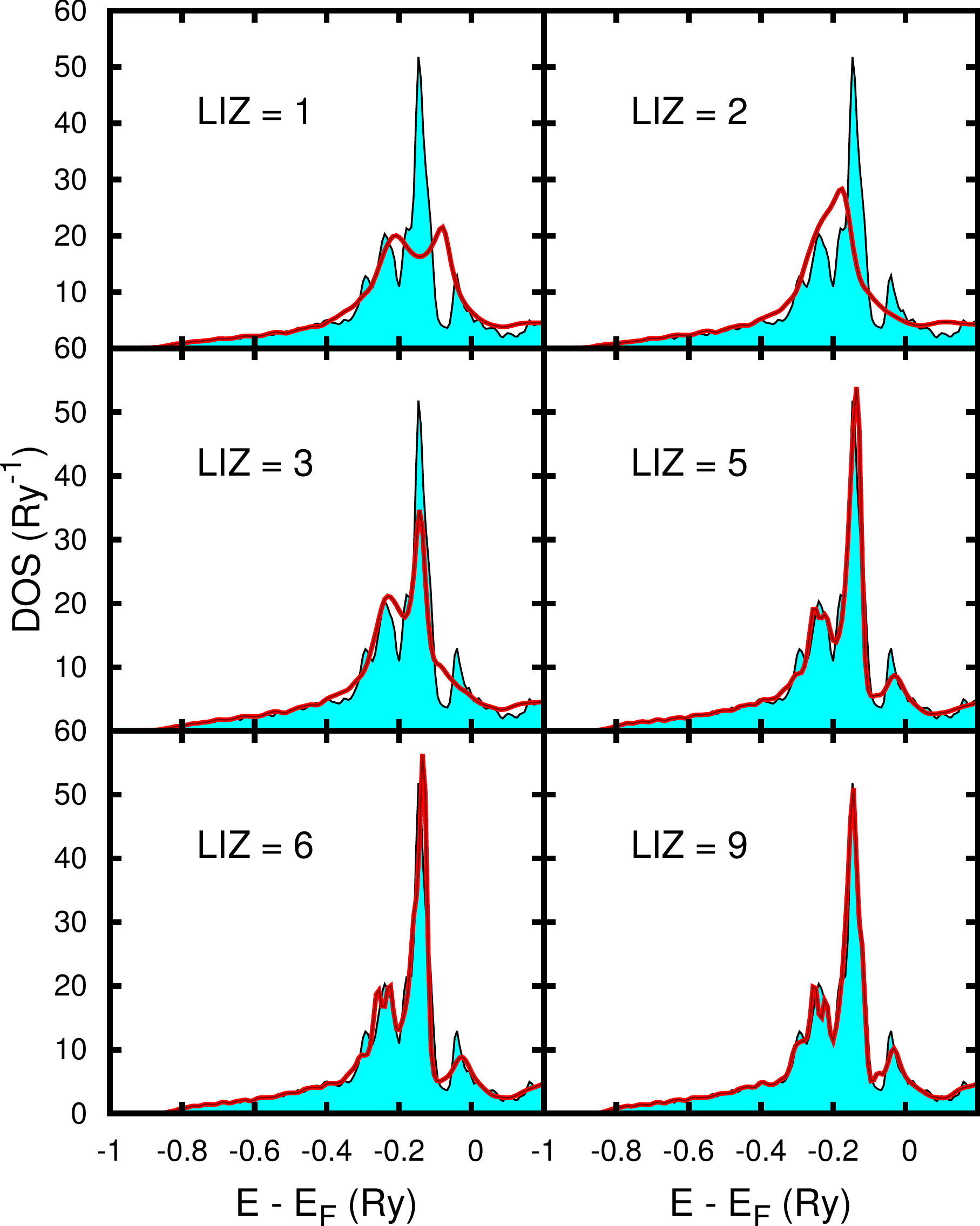}
\caption{(Color online) The DOS for B2-NiAl obtained in the ELSGF
calculations with different LIZ sizes (thick red) compared to
a direct calculation (shaded region). LIZ = 1 corresponds to
a single-site approximation but different from CPA, because
electrostatic potential is calculated for the ordered structure.}
\label{fig:nial_dos}
\end{center}
\end{figure}

%
%
As an example, we present here results for the DOS and the
total energy of a completely ordered B2-NiAl phase, also considered
in Ref.~\onlinecite{abrikosov97}. Let us note that this ordered phase
represents the worst possible case for the LSGF method since its
every coordination shell consists of atoms of only
one type while in a random alloy with the same equiatomic composition,
the number of atoms of both types should be equal (on average).
Formally, this is given by the corresponding atomic correlation
functions, or Warren-Cowley SRO parameters, which take on values
$\alpha_i = -1,1,1,-1,1,1,-1, \dots$ for the first several
coordination shells in the B2 structure. Here, $-1$(1) corresponds 
to the case when every atom
has only atoms of the opposite (same) type at the corresponding
coordination shell, while the SRO parameters are zero in
the case of a random alloy without short-range order effects.

In Fig.~\ref{fig:nial_dos}, we show the DOS of the B2-NiAl phase
obtained in the ELSGF calculations with different sizes of the
LIZ, together with the DOS calculated by the EMTO method.
One can see that at least four coordination shells
(LIZ=5) need to be included in the LIZ to reproduce the main
features of the DOS with a reasonable accuracy, and still some small
DOS features are not well reproduced even when eight coordination
shells (LIZ=9) are included.

%
%

\begin{figure}
\begin{center}
\includegraphics[width=\linewidth]{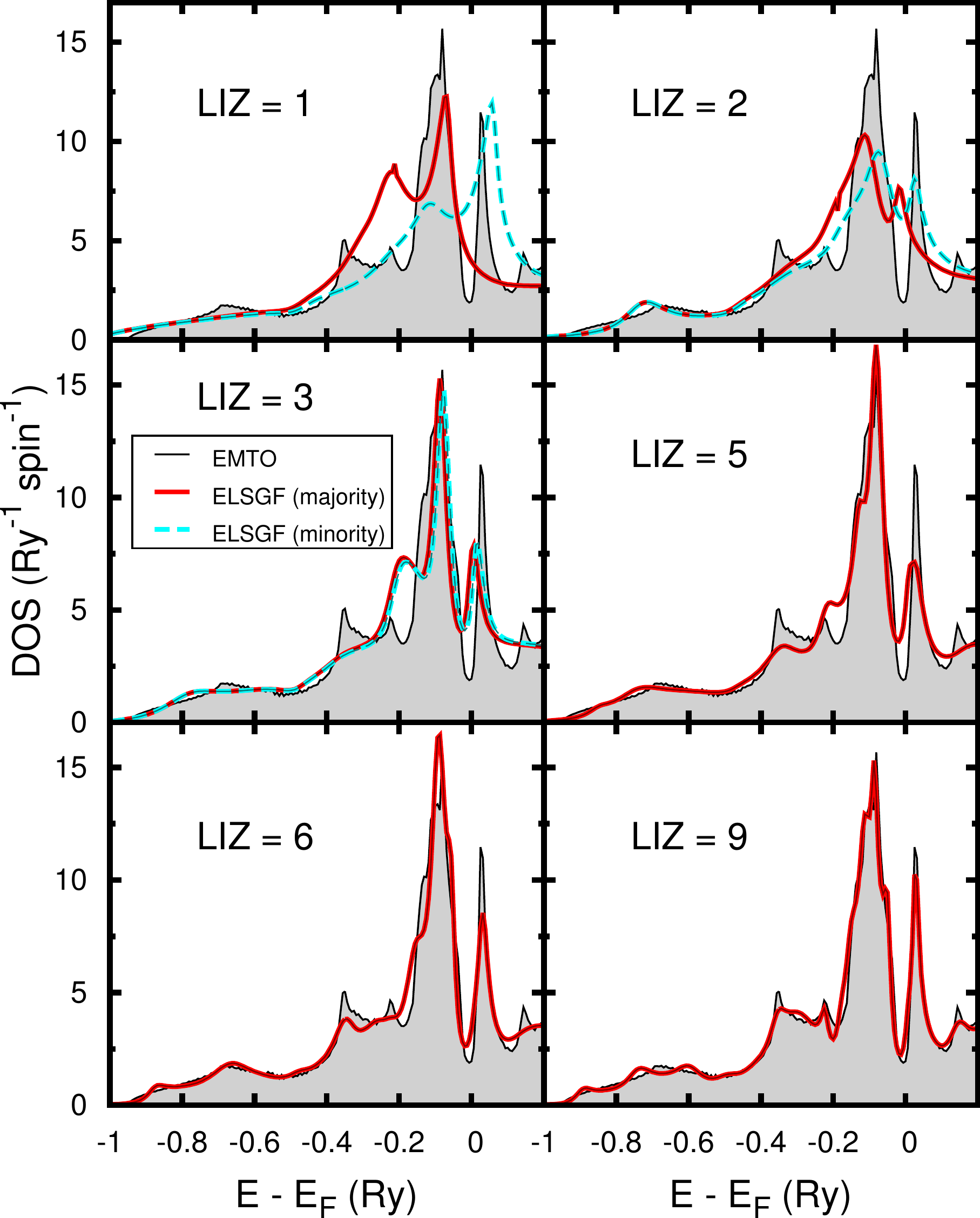}
\caption{(Color online) The DOS for B2-FeSi obtained in the ELSGF
calculations with different LIZ sizes (thick solid-red and dashed-cyan) 
compared to
a direct calculation (shaded region). LIZ = 1 corresponds to
a single-site approximation but different from CPA, because
electrostatic potential is calculated for the ordered structure.}
\label{fig:fesi_dos}
\end{center}
\end{figure}

A similar but a bit more interesting example is shown in
Fig.~\ref{fig:fesi_dos}
where the ELSGF calculations of the DOS of B2-FeSi presented.
This system has unusual magnetic properties very sensitive to
the local environment. In particular, there is quite large
magnetic moment in bcc random Fe$_{0.5}$Si$_{0.5}$ alloy, but
it disappears in the B2 ordered structure.
Performing the ELSGF calculations of the B2 structure with
increasing LIZ one can find the effect of every coordination
shell on the magnetic moment. As one can see in
Fig.~\ref{fig:fesi_dos}, there is an obvious splitting of
the bands in the single-site approximation (LIZ=1,
equivalent to the isomorphous CPA, but with incorrect
electrostatic potential, which is determined for the B2
structure).

\begin{figure}
\begin{center}
\includegraphics[width=\linewidth]{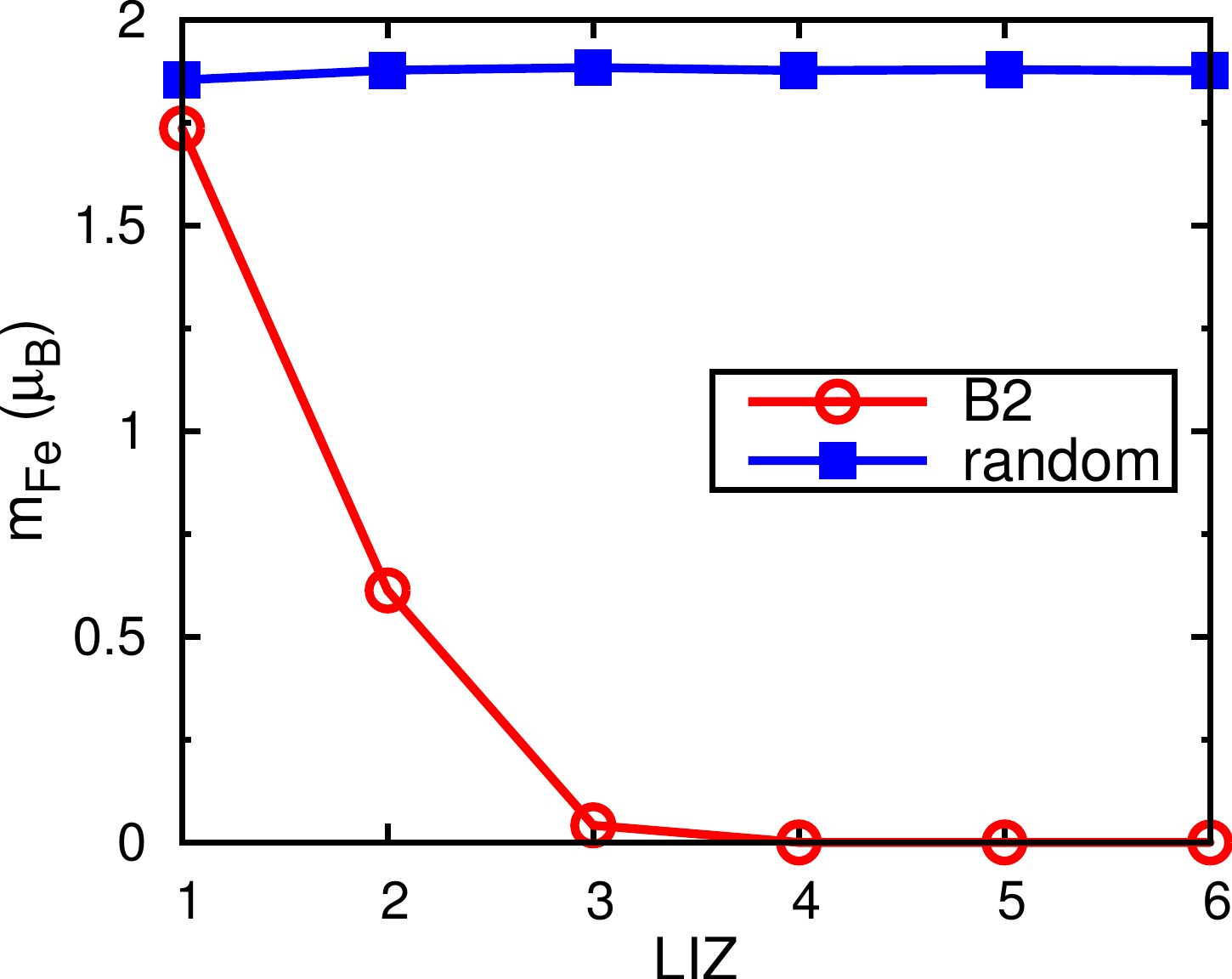}
\caption{(Color online) Local (average over the random alloy) magnetic
moment of Fe in the ELSGF calculations of B2 structure
and an equiatomic random alloy as a function of the LIZ size.}
\label{fig:fesi_m_fm}
\end{center}
\end{figure}

The inclusion of the first coordination shell in the LIZ (LIZ=2),
results in a substantial reduction of the splitting. In this
case Fe atoms can "see" nearest neighbor Al atoms (and vice versa,
Al atoms are surrounded be Fe atoms at the first coordination shell)
while the rest of crystal is represented by the CPA effective medium.
Thus, such a large reduction of the magnetic moment of Fe seems to
be obvious in this case.
The next, second, coordination shell of Fe atoms in the B2 structure
consists of only Fe atoms, and thus one could expect a slight
increase of the magnetic moment due to additional Fe-Fe interactions
when the LIZ size increases from two to three. However, the splitting
and magnetic moment are further reduced, and when the next,
third coordination shell is included in calculations, the 
magnetic moment becomes practically zero (see Fig. \ref{fig:fesi_m_fm}).
This indicates, that most probably the Fe-Fe exchange interaction
parameters are antiferromagnetic in this case. 

\begin{figure}
\begin{center}
\includegraphics[width=\linewidth]{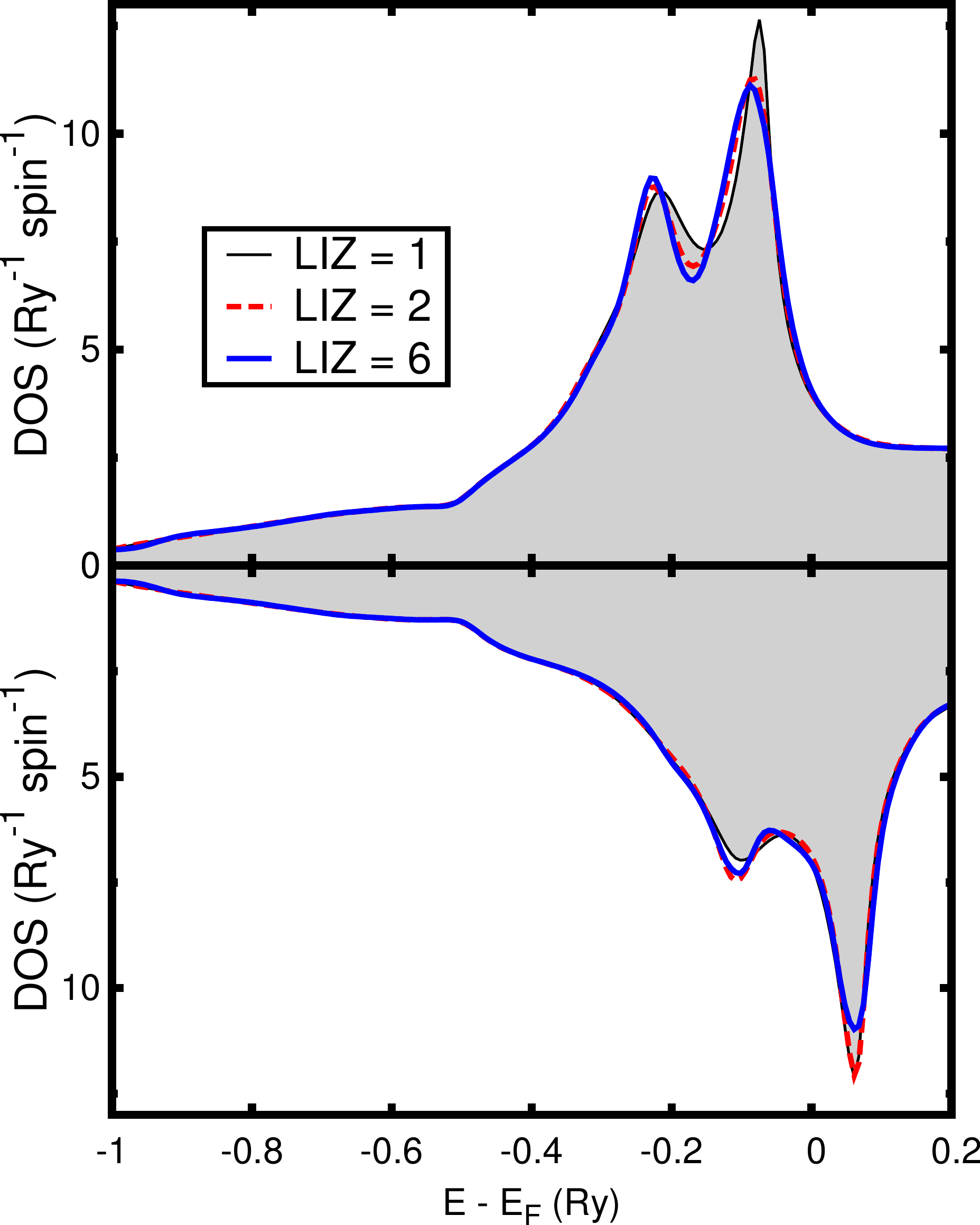}
\caption{(Color online) The DOS of the random bcc Fe$_{0.5}$Si$_{0.5}$
alloy obtained in the ELSGF calculations with different LIZ sizes:
LIZ=1 corresponds to the single-site CPA (shaded area), 
LIZ=2 (thick dashed red line) and 6 (thick solid blue line) are the
results with local environment effects up to the first and
fifth bcc coordination shells, respectively. Spin-majority
band is shown in the upper panel and minority in the 
bottom panel.
}
\label{fig:fesi_rnd_dos}
\end{center}
\end{figure}

For comparison we also show in Fig.~\ref{fig:fesi_rnd_dos}
the DOS of random Fe$_{0.5}$Si$_{0.5}$ alloy calculated by
the ELSGF. In this case the supercell consisting of 256 atoms
($4\times4\times8(\times2)$) atoms has been used, whose atomic
distribution pair correlation functions were as in the random alloy
up to the 8th coordination shell. The DOS for LIZ=1 (isomorphous CPA
model) is actually very close to that for the B2 phase
shown in Fig.~\ref{fig:fesi_dos}. The inclusion of the first
coordination shell in the LIZ leads to a slight modification
of  mostly the spin-majority band, while the inclusion
of more distant coordination shells to a slight change
of the minority spin-band. In Fig.~\ref{fig:fesi_rnd_dos}
we show only the result for LIZ=6 (five coordination shells
included in the calculations). However, this result is
practically indistinguishable from those for the LIZ=4
and 5.

The relatively fast convergence of the DOS with the LIZ
size for random alloys is a natural feature of the LSGF
calculations with the CPA effective medium. The better
the CPA works, the faster convergence. In most cases,
the inclusion just of the first coordination shell in
the LIZ provides very accurate description of the DOS of
random alloys. In the case of Fe-Si it is a bit slow,
exhibiting distant "local environment effect", which are
most probably connected with the non-trivial magnetism
in this system, also showing up in the ELSGF calculations
of the B2-FiSi.

\begin{figure}
\begin{center}
\includegraphics[width=\linewidth]{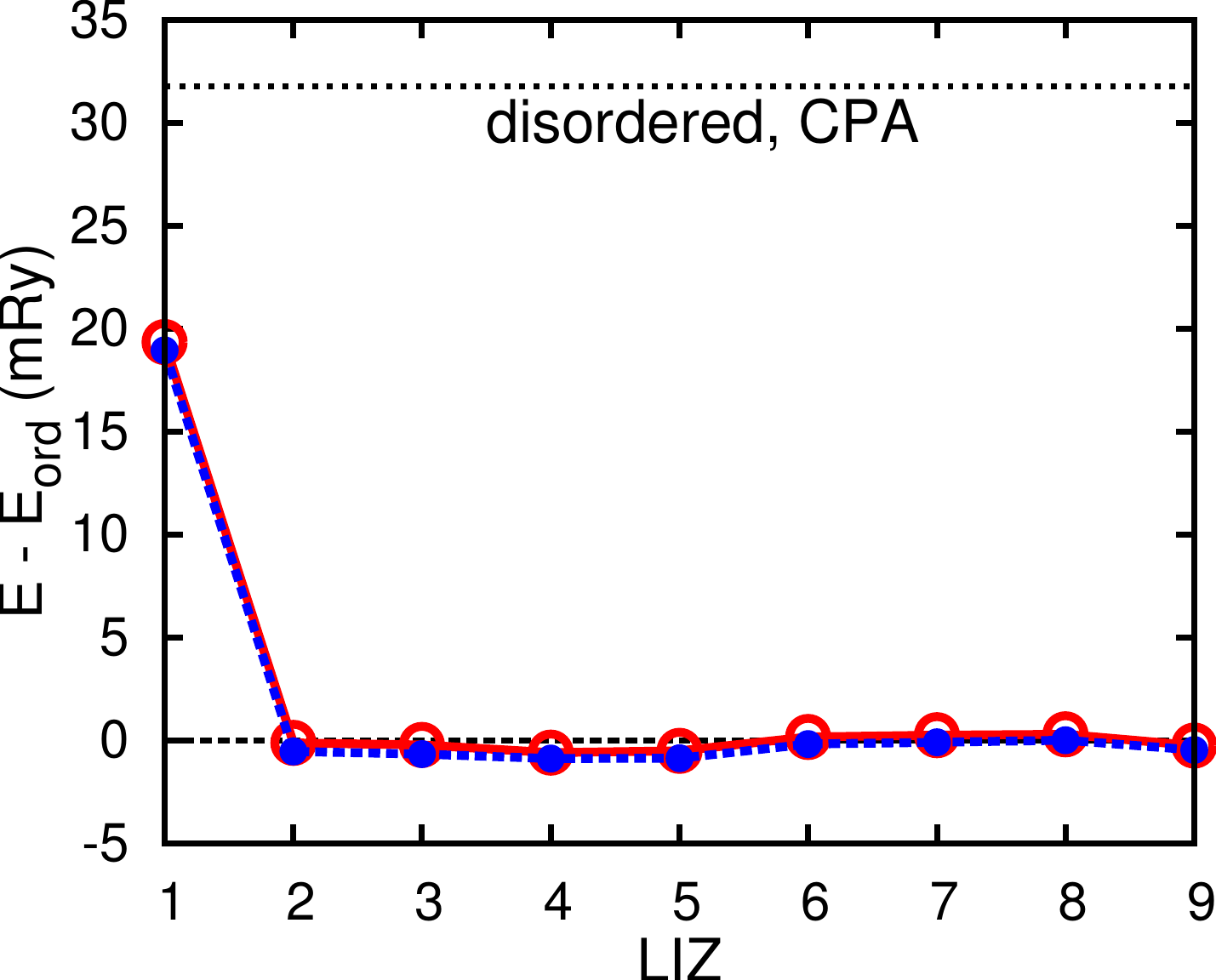}
\caption{(Color online) Convergence of the total energy with respect to
the size of the LIZ for B2-NiAl. The difference between ELSGF (EMTO-CPA -- dotted
line) and EMTO total energies are shown. Red solid line (open circles) --
MT-basis with $l_{\textrm{max}}=3$; blue dashed line (filled circles) --
MT-basis with $l_{\textrm{max}}=2$.}
\label{fig:nial_enr}
\end{center}
\end{figure}

\begin{figure}
\begin{center}
\includegraphics[width=\linewidth]{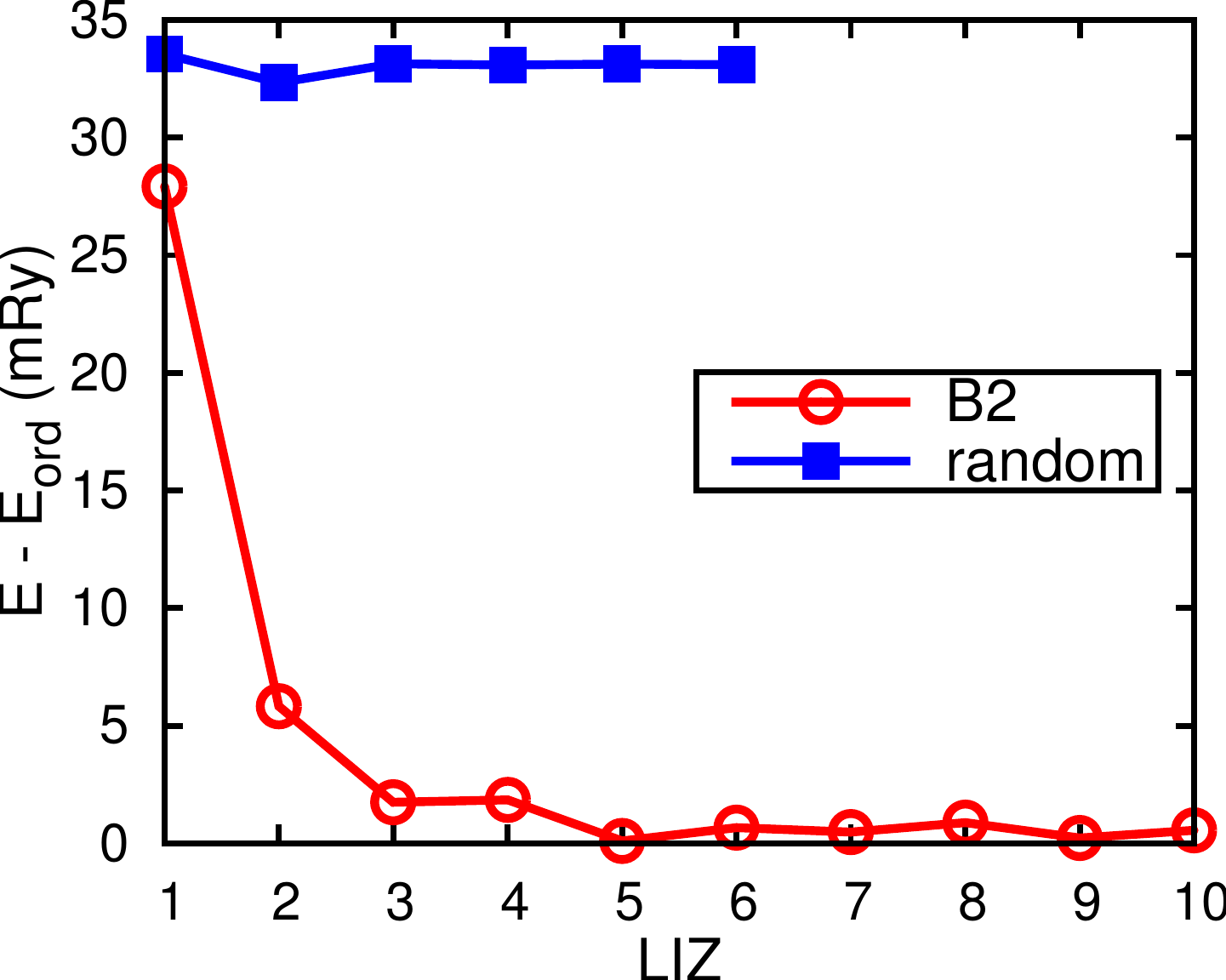}
\caption{(Color online) Convergence of the total energy with respect to
the size of the LIZ for the B2 and random FeSi alloys.}
\label{fig:fesi_enr}
\end{center}
\end{figure}

%
%
The evolution of the density of states with increasing LIZ 
gives an idea of how fast the electronic structure approaches the one
for an ordered system with the inclusion of the corresponding interactions,
thereby providing an estimate of the range of the correlation functions
responsible for specific features of the DOS. An important and
interesting point here, however, is that although the DOS is
directly related to the total energy (specifically, to the band energy),
its details are not important for
the energy value, and the convergence of the total energy with respect to
the LIZ size can thus be much faster. This is partly due to the integral
dependence of the total energy on the DOS. This point
is illustrated in Figs.~\ref{fig:nial_enr} and \ref{fig:fesi_enr},
where we show the convergence
of the total energy of B2 NiAl and FeSi, respectively, 
as a function of the LIZ size.

%
%
One can clearly see that in the case of the B2-NiAl, already starting from
the LIZ corresponding to the first coordination shell (LIZ=2), the difference
in the total energy becomes very small and remains so for larger LIZ.
As was shown in Ref.~\onlinecite{abrikosov97}, the change of the total energy with
the LIZ size can be traced back to the strength of the effective intersite
interactions
entering the Ising configurational Hamiltonian. For this particular
case of NiAl binary alloy, it can be written as
\begin{equation}
H = \frac{1}{2}\sum_p \sum_{i,j \in p} V^{(2)}_p \delta c_i \delta c_j + \dots ,
\label{eq:h_ising}
\end{equation}
where summation runs over coordination shells, $p$, and the corresponding
sites of the lattice, $i$ and $j$; $V^{(2)}_p$ are effective pair
interactions and $\delta c_i = c_i - c$ the concentration fluctuation of
the occupation number, $c_i$, which takes on values 1 and 0,
depending on which alloy component occupies site $i$.

%
%
For the system considered, one can, in fact, estimate the value of
effective interactions. Fig.~\ref{fig:nial_enr} shows that the strongest
interaction is the one related to the first coordination shell. Assuming that
it is the pair effective interaction that is dominant at the first coordination
shell, its value could be assessed as just the difference between the total
energy of a random alloy and that of the ordered one with LIZ=2 (which includes
the first coordination shell), i.e. $V^{(2)}_{1} \sim 32$ mRy
(factor $1/8$ in Eq.~\eqref{eq:h_ising} is canceled exactly by the
bcc coordination number 8). However, one should be aware of the fact
that such a simple estimation is valid only in the case when
the contribution from multisite interactions is small compared
with that from the pair interactions. At the same time, there are
quite a few multisite interactions contributing even in the case of
LIZ=2, when only the first coordination shell is included in the
calculations since they are for the all possible clusters in the
LIZ with one site being the central site of the LIZ. 

%
%
One can also see in Fig.~\ref{fig:nial_enr} that the total energy of
the B2-NiAl for LIZ=1,
which is the single-site approximation, is not equal to that of the random
alloy obtained from the CPA, although LSGF with LIZ=1 is formally equivalent to
the CPA. The reason for the difference is the electrostatic energy, which is
calculated in LSGF for the entire supercell representing an ordered system
in our case. The difference between the LSGF result with LIZ=1 and the energy
of the random alloy can be roughly associated with the electrostatic part of the
effective interactions, or a so-called intersite screened Coulomb interaction.
\footnote{One should note however that strictly speaking, the exact evaluation
of the electrostatic potential and energy implies knowing the exact electron
density of the entire supercell, and the latter is known only approximately
in the LSGF calculations due to the use of the finite LIZ.
For instance, in the case of the B2-NiAl alloy and LIZ=1, the multipole-moment
contributions for a specific coordination of Ni and Al atoms are replaced by
those for the effective medium beyond the first coordination shell
(see Ref. \onlinecite{ruban02b}).
This means that care should be taken to provide a desired atomic distribution
correlation functions in the supercell to avoid a spurious electrostatic
contribution. That being said, the effective electrostatic interactions in
disordered metals are well screened within the first several coordination
shells,\cite{ruban02a,ruban02b} which substantially alleviates the
problem for a sufficiently large supercell containing about 500 atoms
or more.}

In Fig. \ref{fig:fesi_enr} we show the results of similar caculations
but for the ordered-B2 and random FeSi alloys. One can see that
the convergence of the total energy of the B2 phase with respect to the
LIZ size is worse than in the case of B2-NiAl.  Another reason for that,
apart from the convergence of the effective interactions, can be the fact
that FM state is stable up to LIZ=3. The existence of quite long-range
local environment effects in random FeSi alloy have been already
discussed above for the DOS. It also leads to the unusually slow
convergence of the total energy of random alloy, which is as a rule
quite accurate already for the LIZ=2, when the first coordination
shell is included in the LIZ. 

\subsection{Short-range order effects in random FeCr alloys}

%
%
One of the advantages of the LSGF method is the fact that the computational complexity
scales linearly with the number of atoms, which renders possible accurate
first-principles calculations for systems containing up to several thousand
atoms. Such a size of a supercell allows one to model a large variety of alloys
with various concentrations and atomic short-range orders (SRO). In this
section, we demonstrate such a possibility for the case of ferromagnetic Fe-rich
Fe-Cr alloys which have attracted great attention of scientists in different
fields owing to potential applications of these alloys in industry.

%
%
From a scientific point of view, this system is quite interesting and
complicated, when it comes to its accurate first-principles description.
One of the reasons is a complex interplay of magnetism at zero as well as at elevated
temperatures, and interatomic interactions, and consequently,
thermodynamic properties of these
alloys.\cite{olsson03,olsson06,klaver06,ruban08,korzhavyi09}
In particular, the type of alloying abruptly changes with Cr concentration
and temperature\cite{hennion83,mirebeau84,ruban08} in the composition range
of up to about 20 at.\%Cr. While there is a quite strong ordering tendency
between Fe and Cr atoms at low temperatures in the ferromagnetic state and at low Cr
concentration, Fe-Cr alloys exhibit a phase separation behavior at higher
temperatures, close to and above the Curie temperature, and with increasing Cr
content. It was also demonstrated in Ref.~\onlinecite{korzhavyi09} that
the usual Ising model breaks down for this system due to a strong local
environment dependence of the effective interactions.

%
%
It is clear that real Fe-Cr alloys must have a certain amount of atomic
short-range order, and its type and magnitude depends on the thermal treatment
of alloy samples. At the same time, practically all calculations for random
Fe-Cr alloys are done for completely random alloy configurations (in fact,
with rare exceptions, these are just CPA-based calculations). In this section, we
investigate the effect of the atomic short-range order on some of the ground-state
and elastic properties of the Fe-rich FeCr alloys using the ELSGF method.

\begin{figure}
\includegraphics[width=\linewidth]{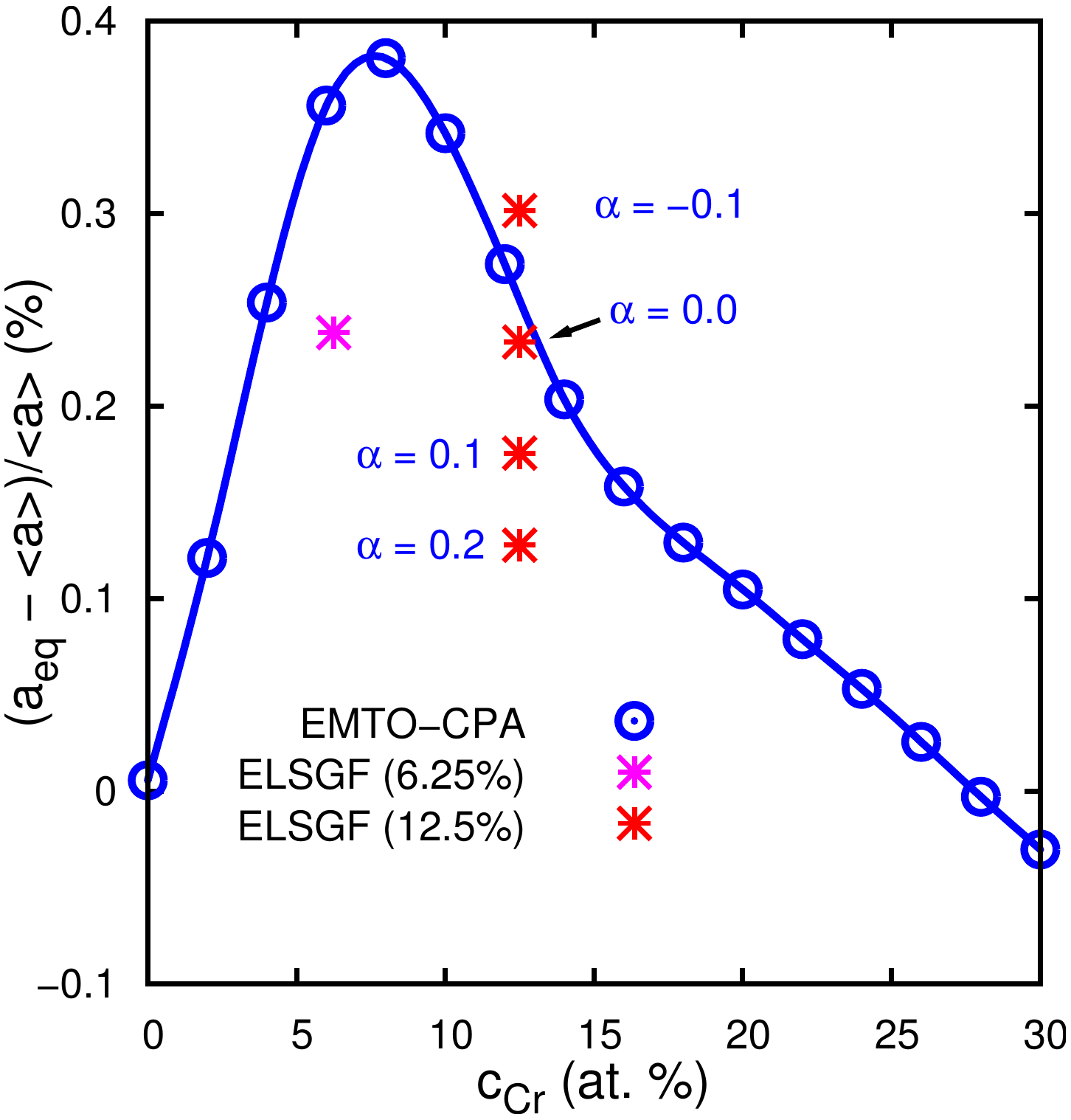}
\caption{(Color online) The dependence of the lattice constant on Cr
concentration, $c_{\rm Cr}$, and short-range order parameter on the first
coordination shell, $\alpha_{1}$, from the EMTO-CPA (circles)
and ELSGF (stars) calculations (the solid line is a guide to the eye).
Deviation of the lattice constant from the Vegard's law is plotted.
The only value obtained with ELSGF for $c_{\rm Cr} = 6.25\%$ is
evaluated for a completely disordered alloy ($\alpha_{i} = 0.0$ up
to the 8th coordination shell).}
\label{fig:latt_const}
\end{figure}

%
%
In Fig.~\ref{fig:latt_const}, we show the generalized-gradient-approximation
(GGA)\cite{perdew96} results obtained by the usual EMTO-CPA method and by the ELSGF for the dependence of the equilibrium theoretical lattice constant on Cr concentration and the (Warren-Cowley) SRO parameter at the first coordination shell,
\begin{equation}
 \alpha_1 = \frac{\xi^{(2)}_{1} - \sigma^2}{1-\sigma^2}.
\end{equation}

The lattice constant is shown in relative units of the deviation from the
average values (given by Vegard's law) for a given concentration.
\footnote{There is a quite strong deviation of the zero-temperature theoretical
results from the room-temperature experimental data for the concentration
dependence of the equilibrium lattice constants in Fe-Cr alloy (see, for
instance, Ref.~\onlinecite{korzhavyi09}). Here, we avoid any discussion of a
possible reason, since it is beyond the subject and the aim of this paper.
This problem will be addressed elsewhere.}
The EMTO-CPA results are similar to those obtained in Ref.~\onlinecite{korzhavyi09} and the ELSGF results have been obtained using a 256-atom supercell (8$\times$4$\times$4($\times2)$ based on the cubic unit cell of the bcc lattice) for two alloy compositions of Fe$_{0.9375}$Cr$_{0.0625}$
and Fe$_{0.875}$Cr$_{0.125}$. In the first case, calculations have been done
only for a completely random alloy without the SRO, while in the latter case, we
have calculated the equilibrium lattice spacing of four supercells with
$\alpha_1 = -0.1$, 0, 0.1, and 0.2. All these supercells are, of course, not
ideally random, but the deviation of the other correlation functions from those
in a random alloy has been small.\footnote{Specifically, for Fe$_{0.9375}$Cr$_{0.0625}$,
the pair SRO parameters have been fixed to 0.0 up to the 8th coordination shell;
in the case of Fe$_{0.875}$Cr$_{0.125}$ for $\alpha_{1} = 0.0$, the first
small non-zero value has appeared at the 7th coordination shell
($\alpha_{7} \approx 0.011$), and for $\alpha_{1} \neq 0.0$, the values at
the other coordination shells have been kept within 30\% of $\alpha_{1}$.}

\begin{figure}
\includegraphics[width=\linewidth]{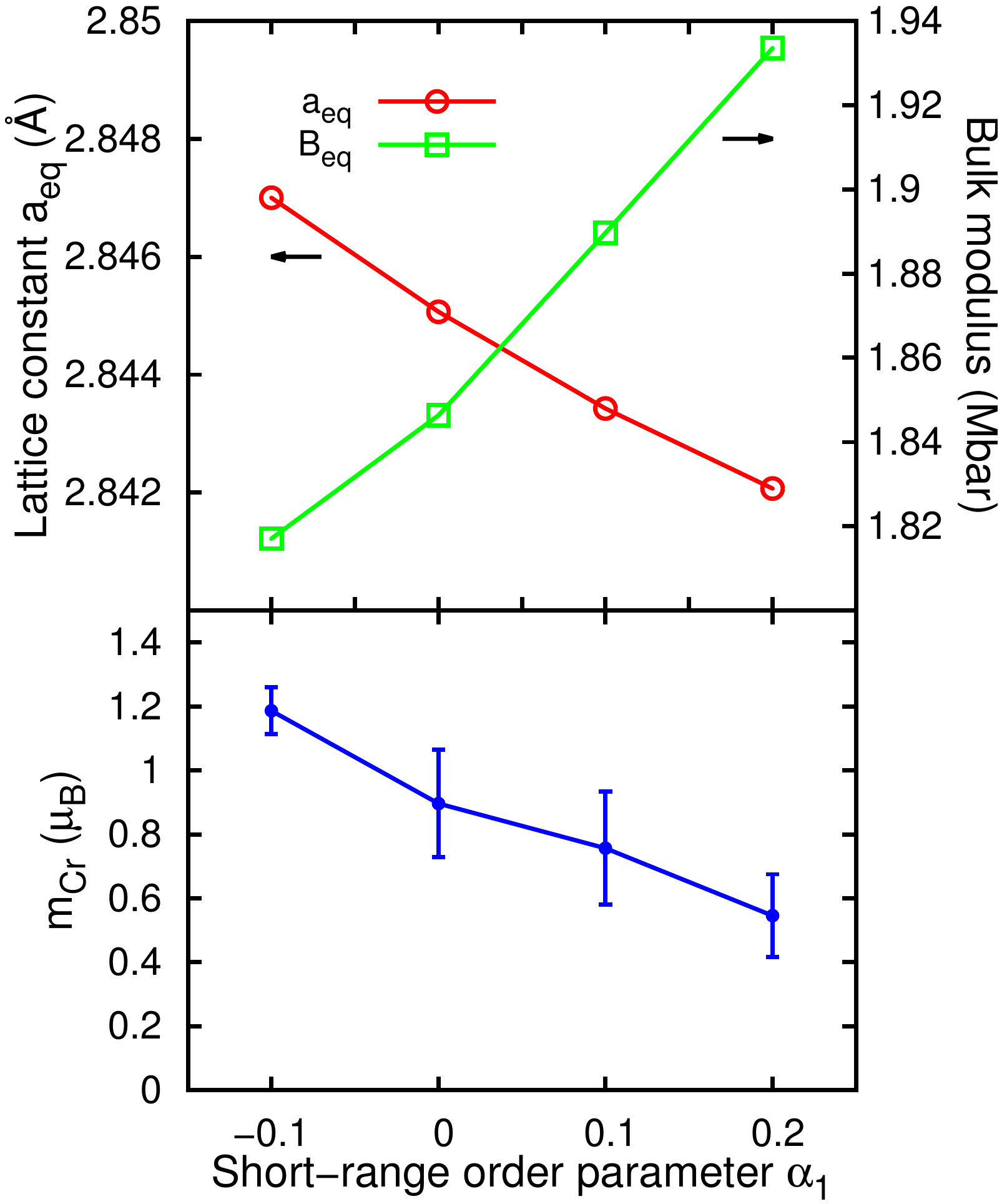}
\caption{(Color online) The dependence of the lattice constant (circles),
bulk modulus (boxes),
and magnetic moment of Cr (dots with whiskers) on the short-range order parameter on the
first coordination shell, $\alpha_{1}$. The whiskers in the bottom
panel show the mean-square-root of the distribution of the
magnitude of magnetic moments on Cr atoms.}
\label{fig:sro_dep}
\end{figure}

%
%
As one can see, the ELSGF results for the random alloy are below the
corresponding CPA results, hence closer to the experimental values.\cite{korzhavyi09}
One of the reasons is that the EMTO-CPA
self-consistent calculations have been performed for fixed values of screening
constants entering the definition of the on-site screened Coulomb interactions
in the single-site DFT formalism.\cite{ruban02a,ruban02b} Although they
were determined for every composition,\cite{korzhavyi09} they were kept
constant in the total energy calculations for different lattice constants.
Local environment effects can also play an important role in this alloy,
since they affect the magnitude of the magnetic moment of Cr atoms dramatically.
\cite{klaver06,ruban08,korzhavyi09} A clear manifestation of this effect is
the dependence of the equilibrium lattice constant in Fe$_{0.875}$Cr$_{0.125}$
on the SRO parameter: it decreases proportionally to the value of $\alpha_1$
(see also the top panel of Fig.~\ref{fig:sro_dep})

%
%
Let us note that the result is counterintuitive, for negative values of the
SRO parameter mean ordering tendency, i.e. preferential occupation of the
corresponding coordination shell by the atoms of the opposite type, while
positive values correspond to phase separation. The reason for such a reversal
is the behavior of the local magnetic moments of Cr atoms. Their magnitude in
Fe-rich Fe-Cr alloys is roughly proportional to the number of Fe nearest
neighbors, and they, thus, go down with increasing the SRO parameter at the
first coordination shell, as one can see at the bottom panel
of Fig.~\ref{fig:sro_dep}. Decreasing magnetic moments of Cr results in
the reduction of magnetic pressure, producing the effect of contraction.
This example shows that SRO effects are especially important in
magnetic alloys.

\subsection{Elastic constants of Fe$_{0.875}$Cr$_{0.125}$}

%
%
The full-charge-density formalism implemented in the EMTO method
allows one to perform relatively accurate calculations of elastic properties
of solids.\cite{vitos07}
Although theoretical results are usually in a quite good agreement with
available experimental data, the accuracy of such calculations, in particular
related to the use of the singe-site CPA in
the DFT self-consistency, is not known.

When no alloys on sublattices are present, the ELSGF and EMTO methods are
equivalent.
In other words, the calculations of ordered systems are equivalent in accuracy,
and this makes possible the direct comparison of the results
on elastic properties obtained by the
EMTO-CPA and ELSGF methods. To this end, we have again chosen Fe-Cr alloys,
whose elastic constants have been recently calculated by the EMTO-CPA method.
In particular, we have calculated shear elastic moduli, $c'$ and $c_{44}$, of
the Fe$_{0.875}$Cr$_{0.125}$ alloy at the experimental lattice constant, 2.869
\AA, in both the ferromagnetic and paramagnetic, i.e. DLM, states.

The ELSGF calculations have been done for a completely random alloy and with
some amount of atomic SRO at the first coordination shell, $\alpha_1=0.1$.
The alloys have been modeled by 512-atom supercells (8$\times$8$\times$8 based on the
primitive unit cell of the bcc lattice). The usual EMTO-CPA calculations have
been done exactly for the same set up of k-points (39$\times$39$\times$39
division of the full Brillouin zone in the Monkhorst-Pack method
\cite{monkhorst76}), $l_{max}$ cut-off (=3 for the partial waves inside atomic
spheres), and other parameters.

For random alloys, an important difference between the EMTO-CPA and ELSGF
calculations is that the DFT-based calculations of
random alloys within single-site CPA should take into consideration
the shift of the one-electron potential due to the on-site screened Coulomb
interactions, $V_i^{scr}$.\cite{ruban02a,ruban02b} As has been
demonstrated in Ref.~\onlinecite{ruban02a,ruban02b},
\begin{equation}
V_i^{scr} = -\alpha_{scr} \frac{e^2 q_i}{S_{ws}} ,
\end{equation}
where $q_i$ is the net charge in the atomic sphere of the $i$-th alloy
component, $S_{ws}$ Wigner-Seitz radius, and $\alpha_{scr}$ the screening
constant which can be determined from the average values of the net charges
and electrostatic potentials of the alloy components in supercell
calculations.\cite{ruban02a,ruban02b} The total energy in this case should be
also corrected by the energy of the on-site screened Coulomb interactions,
$E_{scr}$, which in the case of a binary alloy is
\begin{equation}
E_{scr} = -\frac{e^2}{2}\sum_i c_i \alpha_{scr} \beta \frac{q_i^2}{S_{ws}}.
\end{equation}
Here $c_i$ is the concentration of the $i$-th alloy component, and an
additional coefficient $\beta$ takes care of the non-spherical contributions to
the electrostatic energy ($\beta = 1$ if the multipole moment contributions to the
electrostatic energy and potential are neglected).\cite{ruban02b}

To do accurate single-site DFT-CPA calculations of the total
energy of a random alloy, one should first determine both screening constants,
$\alpha_{scr}$ and $\beta$. This can only be done in supercell calculations,
which enables one to go beyond the single-site approximation and determine the
electrostatic energy and potential accurately. Obviously, such calculations are
computationally very demanding, and the screening constants are, therefore,
usually assumed to be constant for a given alloy, or at most have some
concentration dependence.\cite{korzhavyi09} In our EMTO-CPA calculations here,
we determine screening constants from the corresponding ELSGF
calculations for the initial undistorted bcc structure. In the FM calculations,
we get $\alpha_{scr} = 0.7129$ and in the DLM calculations, $\alpha_{scr}=
0.778$.

The ELSGF calculations have been done with LIZ=3 for the undistorted bcc lattice
(two nearest-neighbor coordination shells for every site) and LIZ=5 for the
distorted lattices. The shear moduli have been determined from fitting
the total energies of the alloy for five distortions with the step 1 and 0.5 \%
in the cases of $c'$ and $c_{44}$, respectively. The results of the calculations
are presented in Table~\ref{tbl:el_const}.

\begin{table}[1th]
\squeezetable \caption{Elastic constants (in Mbar) of the Fe$_{0.875}$Cr$_{0.125}$
random alloy with and without SRO at the first coordination shell
obtained by the EMTO-CPA, ELSGF and PAW methods in the ferromagnetic and DLM
states.}
\begin{center}
\begin{ruledtabular}
\begin{tabular}{c|cc|c|c}
         & \mc{2}{c}{ELSGF}                  & EMTO-CPA     &  PAW     \\
Constant & $\alpha_1 = 0$ & $\alpha_1 = 0.1$     &  ($\alpha_1 = 0$) & ($\alpha_1 = 0$)   \\
\hline
$c_{44}$-FM   & 1.063        &  1.090            & 1.099   &  0.900   \\
$c_{44}$-DLM  & 1.196        &  1.149            & 1.194   &         \\
$c'$-FM       & 0.705        & 0.741             & 0.740   &  0.555   \\
$c'$-DLM      & 0.217        & 0.226             & 0.231   &         \\
\end{tabular}
\end{ruledtabular}
\end{center}
\label{tbl:el_const}
\end{table}%

\begin{figure}
\includegraphics[width=\linewidth]{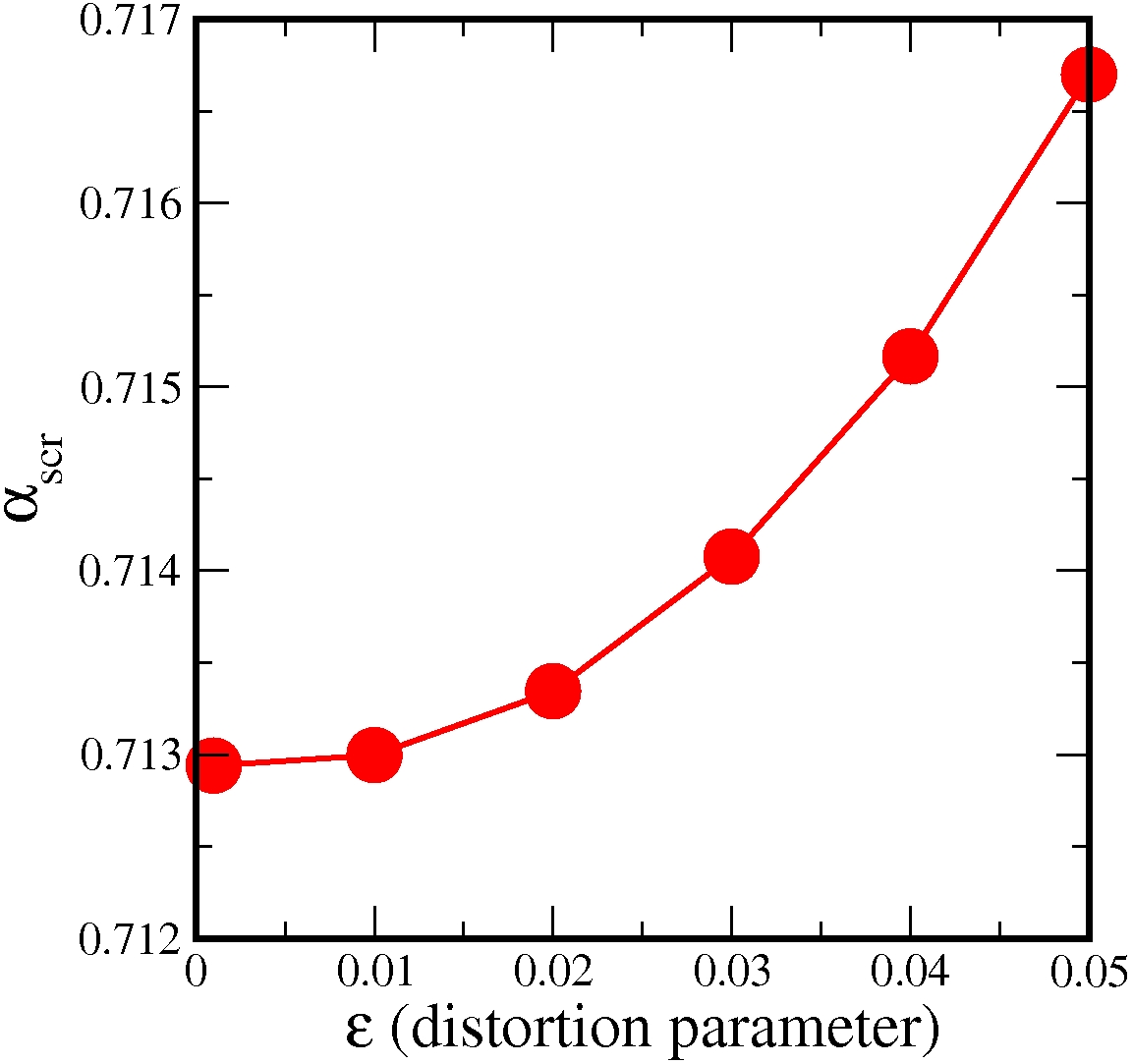}
\caption{The screening parameter, $\alpha_{scr}$, as a function of
the distortion parameter in the calculation of $c'$ in
the ferromagnetic Fe$_{0.875}$Cr$_{0.125}$ random alloy.
}
\label{fig:a_scr}
\end{figure}

As one can see from the table, the EMTO-CPA calculations overestimate (in this particular
case, of course) both elastic constants. In the case of ferromagnetic
calculations, this results from the assumption of the independence of the
screening constants, $\alpha_{scr}$ on the amount of deformation. In
Fig.~\ref{fig:a_scr}, we demonstrate that the assumption is not really
accurate. The screening constants have been determined in the corresponding
ELSGF calculations as
\begin{equation}
\alpha_{scr} = -e^2 S_{ws}\frac{\langle V_i\rangle - \bar{V}}{\langle q_i\rangle},
\end{equation}
where $\langle V_i \rangle$ and $\langle q_i \rangle$ are the average values of the electrostatic potential
and net charge of the atomic sphere of the $i$-th alloy component in the
supercell, and $\bar{V} = \sum_i c_i \langle V_i \rangle$ ($\bar{V} = 0$ in the absence of
the multipole-moment contributions). The screening constant does not depend on
the alloy component in the case of binary alloys, but it becomes
component-dependent if the number of alloy components is greater than two.

The screening constant growing with the deformation parameter implies the decrease
of the screened-Coulomb-interaction energy and as a result, of the total
energy of the alloy. This produces the effect of softening of the elastic
constants. In CPA calculations, this effect is neglected and higher values of
the shear moduli are obtained. It is interesting to note that the
phase-separation type of the atomic SRO leads to the increase of both shear
moduli by about the same 5\% as the error coming from the incorrect
electrostatics in the EMTO-CPA calculations.

There is also a noticeable effect of the magnetic state on elastic properties.
Going over from FM to the DLM state in Fe$_{0.875}$Cr$_{0.125}$ results in a
significant drop of $c'$ by about a factor of 3, and $c_{44}$-constant
increases by about 10\% (more thoroughly this system is discussed in
Ref. \onlinecite{razumovskiy11}). It is clear that the effect of the
atomic SRO on the elastic constants in FeCr alloys in the paramagnetic
state is much less pronounced than in the ferromagnetic case. The difference
between EMTO-CPA and ELSGF calculations is also small in the paramagnetic state.
This weak sensitivity of the strain with respect to the magnetic moment, as well as
the screening constant, is due to isotropy of the system in this case.

In Table~\ref{tbl:el_const}, we also show the results of the
projector-augmented-wave (PAW) \cite{Blochl94,Kresse99} calculations of the
random Fe$_{0.875}$Cr$_{0.125}$ alloy that has been modeled by a 128-atom
supercell.\cite{razumovskiy11} Clearly, the present implementation of
the total-energy full-charge-density technique produces quite a substantial
error, of about 25-30\%, and this issue needs a separate investigation.
Nevertheless, ELSGF seems to be a versatile and sufficiently accurate tool
to investigate complex effects of the atomic SRO, especially
when it comes to the impact of a magnetic state
on various properties of alloys.

\subsection{Surface segregation energies of Cr on Fe(001) and Fe(011)}

%
%
Apart from direct applications of LSGF to random alloys, the method
can also be used to study inhomogeneous systems, such as surfaces,
interfaces, impurities etc.
One of such prospective applications is the calculation of solution
and segregation energies in cases when the size mismatch of alloying components
is small (and the relaxation energy is small). The advantage of the LSGF formalism
here is that the LIZ effectively cuts
off a spurious interaction between alloying species. This property is very
important for systems either with strong and long-range effective
interactions and/or exhibiting strong concentration dependence of the alloying
behavior. This is exactly the case of Fe-Cr alloys, where (as has been
demonstrated in Ref.~\onlinecite{ponomareva07}), the size of the supercell
plays a very important role in the determination of the surface segregation
energy.

An alternative way to calculate this energy is to use the CPA-based method.
\cite{ruban99} However, in the presence of a substantial charge transfer
between atomic spheres of the alloy components, it is very difficult to
accurately take care of the on-site screened Coulomb interactions for surface
alloys, and this can lead to a large error as was demonstrated in
Ref.~\onlinecite{nilekar09}.

\begingroup
\squeezetable
\begin{table}[2th]
\caption{Surface segregation energies (in eV) of Cr on the (100) and
(110) surfaces of Fe in the FM and DLM states. }
\begin{center}
\begin{ruledtabular}
\begin{tabular}{ccc}
               &  Fe(110) &  Fe(100) \\
\hline
FM             & 0.068    &  0.204  \\
DLM            & 0.144    &  0.190  \\
\end{tabular}
\end{ruledtabular}
\end{center}
\label{tbl:segr_en}
\end{table}
\endgroup

In this section, we apply the ELSGF method to the
calculation of the surface segregation energy of Cr onto the (001) and (110)
surfaces of bcc Fe.
Although this energy has nothing to do with the corrosion resistance of
steels, as has been frequently claimed, it is anyway an important thermodynamic
quantity.

There exists several calculations of the surface
segregation energies for this system,
\cite{nonas98,ruban99b,geng03,ponomareva07,kiejna08} but all of them have been
done for the ferromagnetic state.
An additional advantageous feature of LSGF is that the surface segregation energy (or more general
surface-related quantities) can be obtained in
the paramagnetic (DLM) state, which is important because phase
transformations usually take place in this state.

The surface segregation energy is the energy difference between two
configurations of an impurity atom: one with the impurity being in the surface layer and
another one, when it is in the bulk. The supercell approach is then reduced to the
calculations of such two systems. The surface in this case can be modeled using
a slab geometry with a vacuum region which, in the case of the EMTO method, is
filled with empty spheres. In our case, we have chosen a 20-layer slab for the
(001) surface (13 atomic and 7 vacuum layers) and a 14-layer slab (9 atomic and
5 vacuum layers) for the (110) surface. The corresponding 20- and 14-atom unit
cells have been used to define the effective medium, while the entire
supercells have been constructed from the initial slab unit cells by translations
in the plane parallel to the surface layer repeating the unit cells
6$\times$6 and 8$\times$6
times for the (100) and (110) surfaces, respectively. The supercell has
consisted, thus, of 672 sites for the
(110) surface and of 720 sites for the (001) surface.

In the ELSGF calculations, the LIZ has consisted of a central atom and its two
nearest-neighbor (bcc) coordination shells (LIZ=3). The self-consistent
calculations have been done for the room-temperature experimental lattice
constant of Fe, 2.86 \AA, using local-density approximation (LDA).
\cite{perdew92}

The results of the calculations are presented in
Table~\ref{tbl:segr_en}. They can be compared to the results for the (100) surface
obtained by Ponomareva {\it et al.},\cite{ponomareva07} who also did
calculations for the room-temperature lattice constant of Fe using the PAW
method and found that the surface segregation energy of Cr on the (100)
surface of Fe is 0.190 eV. Our ELSGF result, 0.204 eV, is in a good agreement
with the PAW result.

The segregation energies obtained in Ref.~\onlinecite{kiejna08} are
significantly lower than those in Table~\ref{tbl:segr_en}: -0.001 and 0.076 eV
for the (110) and (100) surfaces, respectively, most probably because they were obtained for a
relatively small supercell and for the theoretical equilibrium lattice constant
of Fe. As was demonstrated by Ponomareva {\it et al.},\cite{ponomareva07}
both these parameters strongly affect the results, and we therefore believe that
our results are quantitatively accurate.

Finally, one can see that there is quite a pronounced effect of the magnetic
state on the segregation energy of the (110) surface: the surface segregation
energy in the DLM (paramagnetic) state is almost doubled compared to that in
the FM state. This means that at elevated temperatures, relevant to experimentally
achievable equilibrium, the surface segregation of Cr atoms towards the (110)
should be reduced. Let us note that this surface has an important role in the
thermodynamics since it is the most closely packed surface in the case of the
bcc structure, and thus has the lowest energy. It is also clear that such
a result is extremely difficult, if possible, to obtain at the same level of
accuracy by any other existing method.

%
%
\section{Conclusion}

%
%
We have demonstrated that the ELSGF method can be a rather accurate
and versatile tool for studying local-environment effects in random alloys
as well as in inhomogeneous systems, such as surfaces and interfaces.
In particular, we have applied it to the Fe-rich FeCr alloy and found that
an intimate coupling between SRO and the equilibrium lattice constant as well as
elastic properties results from the strong sensitivity of Cr magnetic moments
on the local environment.
Among the advantages of the method is its order-$N$ scaling, which makes
the implementation easy to parallelize and enables one to treat
large supercells consisting of $N \sim 10^{4}-10^{5}$ atoms.
In addition, the capability
to treat the high-temperature paramagnetic state renders possible
investigating phase transitions in magnetic alloys.

%
%
Compared to the KKR-ASA implementation, ELSGF has a much more
accurate normalization of states and can, therefore, be applied
to systems with a distorted structure and large ion-size mismatches.
This feature opens up a completely new possibility to take into account
random relaxations in alloys within ELSGF. Formally, this amounts to introducing
an additional perturbation of the structure constants into the
Dyson equation, Eq.~\eqref{eq:ecm}. Accurate evaluation of
the kinetic energy is a necessary prerequisite for such an expansion
in the structure-constant perturbation to give good results, which can be achieved
by a full-charge-density-self-consistent implementation of ELSGF.

Other implementations of LSGF are possible. For instance,
a fully-relativistic EMTO-LSGF method can be realized
by replacing Green's functions in spin-up-spin-down channels with
full Dirac $2\times2$-spinor Green's functions. A full-potential-KKR
implementation is also straightforward.

\begin{acknowledgments}
The authors would like to thank I.A. Abrikosov and L.~Vitos for
discussions. OEP expresses gratitude to A.~Lichtenstein
for useful comments. AVR and BJ acknowledge the financial support
of the support of the Swedish Research Council (VR) and European
Research Council (ERC). Computer resources for this study have been
provided by the Swedish National Infrastructure for Computing (SNIC)
and MATTER Network, at the National Supercomputer Center (NSC),
Link\"{o}ping. This work was partly performed within the VINNEX
center Hero-m, financed by the  Swedish Governmental Agency for
Innovation Systems (VINNOVA), Swedish industry, and the Royal
Institute of Technology (KTH).
\end{acknowledgments}

\bibliography{ref_elsgf}

\end{document}